\documentclass[12pt]{article}
\usepackage{amsfonts,amsmath,amssymb,epsf,cite}
\usepackage{graphicx,color,subfigure}
\usepackage[usenames,dvipsnames]{pstricks}
 \def\a{{\alpha}}
 
 \def\frac#1#2{{#1\over #2}}

\def\cA{{\cal A}}
\def\cB{{\cal B}}
\def\cJ{{\cal J}}

\def\nn{\nonumber \\}

 \def\a{{\alpha}}
 
 \def\frac#1#2{{#1\over #2}}

 \def\cO{{\cal O}}

 \def\ben{\begin{equation}}
\def\een{\end{equation}}
\def\beq{\begin{eqnarray}}
\def\eeq{\end{eqnarray}}
\def\bea{\begin{eqnarray}}
\def\eea{\end{eqnarray}}

\begin{document}
\begin{titlepage}
\thispagestyle{empty}
\begin{flushright}
UK/14-05\\
\end{flushright}

\bigskip

\begin{center}
\noindent{\Large \textbf
{Kibble-Zurek Scaling in Holographic Quantum Quench : Backreaction}}\\
\vspace{2cm} \noindent{
Sumit R. Das $^{(a,b)} $\footnote{e-mail:das@pa.uky.edu} and
Takeshi Morita $^{(a,c)} $\footnote{e-mail:morita.takeshi@shizuoka.ac.jp}

\vspace{1cm}
 $^{(a)} ${\it Department of Physics and Astronomy, \\
 University of Kentucky, Lexington, KY 40506, USA} \\
$^{(b)}$ {\it Yukawa Institute for Theoretical Physics, \\
Kyoto University, Kyoto 606-8502, JAPAN} \\
$^{(c)}$ {\it Department of Physics,\\
Shizuoka University, 
836 Ohya, Suruga-ku, Shizuoka 422-8529, JAPAN}}

\end{center}

\vspace{0.3cm}
\begin{abstract}
We study gauge and gravity backreaction in a holographic model of
quantum quench across a superfluid critical transition. 
The model involves a complex scalar field coupled to a gauge and gravity field in the bulk. In earlier work (arXiv:1211.7076) the scalar field had a strong self-coupling, in which case the backreaction on
both the metric and the gauge field can be ignored. In this
approximation, it was shown that when a time dependent source for the
order parameter drives the system across the critical point at a rate
slow compared to the initial gap, the dynamics in the critical region
is dominated by a zero mode of the bulk scalar, leading to a Kibble-Zurek
type scaling function. We show that this mechanism for emergence of
scaling behavior continues to hold without any self-coupling in the
presence of backreaction of gauge field and gravity. Even though
there are no zero modes for the metric and the gauge field, the scalar
dynamics induces adiabaticity breakdown leading to scaling. This yields scaling behavior for the time dependence of the charge density and energy momentum
tensor.

\end{abstract}
\end{titlepage}
\newpage

\tableofcontents
\newpage

\section{Introduction and summary}

Quantum (or thermal) quench across critical points is an interesting
problem in many areas of physics. Consider starting in the gapped
phase of a system and turning on a time dependent external parameter
which drives it to a critical point at a rate {\em slow} compared to the
initial gap. While the initial time evolution will be adiabatic,
adiabaticity will break down close to the critical point and the
subsequent time evolution is expected to carry universal signatures of
the critical point. Many years ago, Kibble \cite{Kibble:1976sj}, and
subsequently Zurek \cite{Zurek:1985qw}, argued that observables like defect
density indeed show scaling behavior. These arguments - which were
first developed for thermal quench and recently generalized to quantum
quench \cite{more} \cite{sondhi} - imply that for a
driving involving a single relevant operator, the time dependence
of the one point function of an operator $\cO$ with conformal
dimension $x$ is of the form \cite{dzia}
\ben \cO(t,v) \sim
v^{\frac{x\nu}{z\nu+1}} F(tv^{\frac{z\nu}{z\nu+1}})
\label{0-1}
\een
where $v$ is the rate of change of the coupling, $\nu$ is the
correlation length exponent and $z$ is the dynamical critical
exponent. The arguments which lead to (\ref{0-1}) involve (i) an
assumption that once adiabaticity breaks the system evolve in a 
diabatic fashion and (ii) in the critical region the instantaneous
correlation length is the only length scale in the problem. The first
assumption is rather drastic. The second assumption is reasonable, but
unlike equilibrium critical behavior there is really no well understood
conceptual framework like the renormalization
group which explains why all other scales decouple from the
problem. This is particularly so for strongly coupled systems.
 Nevertheless, Kibble-Zurek scaling has been verified by explicit
calculations in many models and is now being seen experimentally as
well \cite{more,more2}.

In \cite{basu-das} a study of this problem in strongly coupled field theories which have gravity duals via AdS/CFT was initiated and continued in \cite{bddn} and \cite{bdds}. The idea is to use holographic techniques to investigate scaling behavior for slow quench {\em without making any of the above assumptions}.
 In the AdS/CFT correspondence a time dependent coupling of
a strongly coupled boundary field theory corresponds to a time
dependent boundary condition for the bulk dual field, so that the
problem reduces to differential equations with time dependent boundary
conditions. A mechanism for emergence of scaling
emerges in these studies. These models involve bulk scalar fields
which are dual to order parameters and the critical point is
characterized by a zero mode of the scalar, i.e. a solution of the
linearized equations of motion which satisfy zero source boundary
conditions at the AdS boundary and regularity in the interior. It
turns out that in the critical region where adiabaticity is broken (so that
a Taylor expansion in $v$ breaks down) , there is a new small-$v$
expansion in {\em fractional} powers of $v$. To leading order in this
expansion, the dynamics is dominated by the zero mode, and the
resulting bulk equations of the zero mode lead to a scaling solution. The analysis can be also used to determine the corrections to scaling.

These studies did not include the effect of fluctuations (i.e. 1/N
corrections in the boundary field theory). More recently
\cite{Sonner:2014tca} have studied the problem by modelling these
fluctuations with a noise in the time evolution in a manner consistent
with the fluctuation-dissipation theorem and found consistency with
the Kibble Zurek mechanism. Other aspects of quantum quench which involve critical points have been investigated in \cite{otheraspects}, \cite{Buchel}.

The models considered in \cite{basu-das} and \cite{bddn} have scalar
fields in the bulk with strong self-couplings, together with gravity and a Maxwell field. The strong self-coupling allows a probe
approximation in which the backreaction of both the gravity and the
bulk gauge field can be ignored, as in \cite{liu1}. It is important to
examine the effects of backreaction. First, as we will see below, the
zero mode is present only in the scalar sector - not for the gauge
field or the metric perturbations. It is therefore of interest to know
whether the critical dynamics of the gauge field and metric also
simplifies and lead to scaling properties of the charge density and energy-momentum tensor in the boundary theory.
Perhaps more importantly, it is interesting to know
whether the system thermalizes in any sense at late times. This
requires a complete treatment of the dynamics of the bulk metric.  For
a slow driving far away from any critical point, the evolution is
essentially adiabatic. If we start from the ground state, as in the
zero temperature cases of \cite{bddn} and \cite{bdds}, this means that
there is no collapse into a black hole. If the quench crosses a
critical point, the system gets excited and it would be interesting to
know what happens in the bulk.

In this paper, we take the first step in incorporating backreaction
by addressing the first question above. We will find that even though
there is no zero mode in the gauge and gravity sectors, the scalar
zero mode feeds in through nonlinearities 
and leads to a breakdown of adiabatic evolution of
the gauge and gravity fields.  In the critical region there is again
an expansion in fractional powers of $v$.  The scalar dynamics is
dominated by the zero mode and the nonlinear coupling with the gauge
field and the metric leads to scaling solutions for all the
fields. The AdS/CFT dictionary then yields scaling functions for the
expectation value of the order parameter, the charge density and the
energy momentum tensor. 
We will not address the question of late time behavior and
thermalization : this would require detailed numerical work which we
postpone to a later investigation.  

For this purpose, we consider the holographic superfluid model of \cite{rnt}, subsequently studied by \cite{horo}. The model considered in \cite{bddn} is a variation of this model : the scalar has a self-coupling in addition to minimal coupling to the gauge field and the metric. One of the boundary space directions is compact with some radius $R$. Quantum quench is performed by introducing a
time dependent boundary condition which corresponds to a time
dependent source for the order parameter in the boundary field
theory. In \cite{bddn} non-linearity arose from self coupling of the
scalar. In this paper, however, we set the self-coupling to zero - as
in the work of \cite{rnt}.

Now the backreaction of the gauge field cannot be ignored. However,
when the charge of the field is large, there is a probe approximation
where the backreaction of gravity can be ignored (which was used in
\cite{rnt}). We first consider this probe approximation. We determine the equilibrium exponents, and then proceed to examine
the breakdown of the adiabatic expansion. We show that the zero mode
of the scalar field leads to a breakdown of adiabaticity in {\em both}
the scalar and the gauge sector. The time of breakdown is the same for
both the fields - this serves as a consistency check on the
calculation. We then examine the dynamics in the critical region
closely following \cite{basu-das} - \cite{bdds}. In a way analogous to
these works we find that there is a consistent small-$v$ expansion in
fractional powers of $v$. To leading order of this expansion, the zero
mode of the scalar dominates the dynamics. While there is no zero mode
for the gauge field, the equations of motion determine the dependence
of the gauge field in the AdS radial direction in terms of the scalar
zero mode, which leads again to decoupling of modes. The resulting
leading order dynamics then exhibits scaling behavior like (\ref{0-1}),
and the expansion in fractional powers of $v$ provides a way to
calculate the corrections to scaling.

We then proceed beyond the probe approximation and consider the
backreaction of the metric and show the breakdown of adiabaticity, the
existence of a small-$v$ expansion in fractional powers of $v$ and the
emergence of scaling solutions are quite similar to the gauge field
case.

In Section (\ref{sec2}) we describe the basic setup. Section (\ref{sec3}) deals with the quench dynamics in the probe approximation. In section (\ref{sec4}) we incorporate the backreaction of gravity. Section (\ref{sec5}) contains conclusions and discussions.

\section{The Basic Setup}\label{sec2}
\setcounter{equation}{0}

The bulk action in $d+2$ dimensions is given by
\ben
S = \int d^{d+2}x {\sqrt{g}} \left[ \frac{1}{2\kappa^2} \left( R +
  \frac{d(d+1)}{L^2}\right) - \frac{1}{4}F_{\mu\nu}F^{\mu\nu} -
   \left( |\partial_\mu \Phi - iqA_\mu \Phi|^2 - m^2
  |\Phi|^2 \right) \right] \ ,
\label{1-1}
\een
where $\Phi$ is a complex scalar field with charge $q$ and $A_\mu$ is an abelian gauge
field, and the other notations are standard. Henceforth we will use $L=1$
units. One of the spatial
directions, which we will denote by $\theta$ will be considered to be
compact. The radial direction will be denoted by $r$. The mass of the scalar is chosen in the range
\ben
m^2_{BF} < m^2 < m^2_{BF}+1,
\label{1-1a}
\een
where $m_{BF}^2 = -(d+1)^2/2$ is the Breitenholer-Freedman bound.

The boundary theory has a finite chemical potential $\mu$, so that
\ben
{\rm Lim}_{r \rightarrow \infty}(A_t)\rightarrow \mu .
\label{1-2}
\een
The temperature vanishes.

Let us first set $\Phi=0$ (which is always a solution).  As shown in
\cite{rnt}, there is a value of the chemical potential $\mu = \mu_0$
such that for $\mu < \mu_0$ the preferred  solution to Einstein equation is an
AdS soliton \bea ds^2 & = & \frac{dr^2}{r^2 h(r)} + r^2 \left( -dt^2 +
\sum_{i=1}^{d-1} dx_i^2 \right) + r^2 h(r) d\theta^2 \ ,\nn \\ h(r) &
= & 1 - \left( \frac{r_0}{r} \right)^{d+1} \ , \nn \\ A_t & = & \mu
\ ,
\label{1-3}
\eea
with constant parameters $\mu$ and $r_0$. The periodicity of $\theta$ in
this solution is
\ben
\theta \sim \theta + \frac{4\pi}{(d+1)r_0} \ ,
\label{1-4}
\een
$\mu_0$ is given by
\ben
\mu_{0} = \frac{r_0 (d+1)
  (2d)^{\frac{d-1}{2(d+1)}}}{(d-1)^{\frac{d}{d+1}}(d+1)^{1/2}} \ .
\label{1-5}
\een
For $\mu > \mu_0$ the preferred background is an extremal black brane. We will consider the soliton phase. 

In the remainder of the paper we will rescale all distances to set $r_0 = 1$.

\section{The Probe Approximation}\label{sec3}

We now consider the effect of the scalar field.
In this section we consider the regime \ben q^2 \gg \kappa^2,  \een 
so
that the gravity background can be considered to be fixed. 
(Hereafter we fix $q=1$.)
Gravity
backreaction is treated in the next section.
Thus the spacetime background remains an AdS soliton.  Note that this probe approximation is not the same as the probe approximation of \cite{bddn}. In the latter paper there was a strong coupling of the scalar which allowed a different probe approximation where the backreaction of both the gravity and the gauge field can be ignored.

We will consider configurations which are functions of $t$ and $r$ only, and work in a gauge $A_r = 0$. Translation invariance then implies that we can choose $A_i = 0$ where $i$ denotes the boundary spatial directions, which is consistent with the spatial components of the Maxwell equations.  Then the only non-vanishing gauge field component is $A_t$ which we denote simply by $\phi(t,r)$. The coupled scalar-Maxwell equations then become
\begin{align}
-  \frac{1}{r^2} \left( \ddot \Phi_R + \dot \phi \Phi_I + 2 \phi \dot \Phi_I -\phi^2 \Phi_R \right) + 
\frac{1}{r^d} \partial_r \left( r^{d+2} h \partial_r  \Phi_R \right)
 -m^2 \Phi_R = 0,
\label{eom_phi_R_ads}
 \\
-  \frac{1}{r^2} \left( \ddot \Phi_I - \dot \phi \Phi_R - 2 \phi \dot \Phi_R -\phi^2 \Phi_I \right) + 
\frac{1}{r^d} \partial_r \left( r^{d+2} h \partial_r  \Phi_I \right)
 -m^2 \Phi_I = 0,
\label{eom_phi_I_ads}
 \\
\dot \phi'= 2 r^2 \left( \Phi_I \Phi_R' - \Phi_I' \Phi_R   \right),
\label{eom_ar_ads} \\
\frac{1}{r^d} \partial_r \left( r^{d} h \partial_r  \phi \right)
=\frac{2}{r^2} \left( \Phi_I \dot \Phi_R -  \dot \Phi_I \Phi_R  +\phi |\Phi|^2  \right).
\label{eom_at_ads} 
\end{align}
Here $\cdot$ and $'$ are $t$ and $r$ derivative respectively and the complex scalar field is written in terms of its real and imaginary parts
\ben
\Phi = \Phi_R + i \Phi_I.
\een

Near the AdS boundary $r \rightarrow \infty$, these fields satisfy the asymptotic conditions
\begin{align}
\Phi & \rightarrow J(t) r^{-\Delta_-} \left[ 1+ \cdots \right]+
\chi(t) r^{-\Delta_+} \left[ 1+ \cdots \right],
\label{asymptotics-phi}
 \\
\phi & \rightarrow \mu   \left[ 1+  \cdots \right]-
\rho(t) r^{1-d} \left[ 1+ \cdots \right] ,
\label{asymptotics}
\end{align}
where
\ben
\Delta_{\pm} = \frac{d+1}{2} \pm \sqrt{m^2 + \frac{(d+1)^2}{4}} .
\een

Near the tip of the soliton $r=1$, we write $r = 1+x$ with $x \ll 1$ and the equations of motion become
\begin{align}
-  \left( \ddot \Phi_R + \dot \phi \Phi_I + 2 \phi \dot \Phi_I -\phi^2 \Phi_R \right)  
-(d+1) \partial_x \left(  x \partial_x  \Phi_R \right)
 -m^2 \Phi_R = 0
 \\
-  \left( \ddot \Phi_I - \dot \phi \Phi_R - 2 \phi \dot \Phi_R -\phi^2 \Phi_I \right)  
- (d+1) \partial_x \left( x \partial_x  \Phi_I \right)
 -m^2 \Phi_I = 0
 \\
\dot \phi'= 2  \left( \Phi_I \Phi_R' - \Phi_I' \Phi_R   \right)
 \\
-(d+1) \partial_x \left(x \partial_x  \phi \right)
=2 \left( \Phi_I \dot \Phi_R -  \dot \Phi_I \Phi_R  +\phi |\Phi|^2  \right)
\end{align}
Thus the fields behave as
\begin{align}
\Phi_{R,I} & \rightarrow c^{(R,I)}_1(t)+c^{(R,I)}_2(t) \log x
 \\
\phi & \rightarrow d_1(t)+d_2(t) \log x 
\end{align}
Thus regularity at the tip requires $c_2^{(R,I)} (t) = d_2 (t) =0$.

Among these equations, the equation (\ref{eom_ar_ads}) is a constraint equation associated with our gauge choice. Once this equation is imposed on some constant $r$ slice, the other three equations guarantee that it holds everywhere. Let us therefore impose this at a large $r = R_0$ slice.  Using the asymptotic conditions (\ref{asymptotics}), The left hand side of (\ref{eom_ar_ads}) becomes 
\begin{align}
\dot \phi' & \rightarrow -
(1-d) \dot \rho(t) r^{-d} \left[ 1+ \cdots \right] .
\end{align}
On the right hand side, the power of $r$ of each term is not $r^{-d}$. However there are cancellations,
\begin{align}
2 r^2 \left( \Phi_I \Phi_R' - \Phi_I' \Phi_R   \right) \to &
2 r^{2} \left(  J_I r^{-\Delta_-} +
\chi_I r^{-\Delta_+}  \right)\left( -\Delta_- J_R r^{-\Delta_--1} -\Delta_+
\chi_R r^{-\Delta_+-1}  \right)\nonumber \\
&-
2 r^{2} \left( -\Delta_- J_I r^{-\Delta_--1} -\Delta_+
\chi_I r^{-\Delta_+-1}  \right)\left(  J_R r^{-\Delta_-} +
\chi_R r^{-\Delta_+}  \right) \nonumber \\
=& 2r^{-(\Delta_++\Delta_-)+1} \left( \Delta_+ - \Delta_- \right) \left(
J_R \chi_I-J_I \chi_R 
\right).
\end{align}
Here $J=J_R+i J_I$ and $\chi=\chi_R+i \chi_I$.
Since $\Delta_++\Delta_-=d+1$, the power of $r$ is indeed $r^{-d}$ and we get 
\begin{align}
 \dot \rho(t) = \frac{2 \left( \Delta_+ - \Delta_- \right)}{d-1} \left(
J_R \chi_I-J_I \chi_R 
\right).
\end{align}
This may be finally re-written as
\ben
\partial_t \rho = \frac{2(\Delta_+-\Delta_-)}{d-1} {\rm Im} \left[ J^\star (t) \chi (t) \right].
\label{constrainteqn}
\een

\subsection{The equilibrium critical point and its exponents}

The AdS soliton with constant $\phi = \mu$ is not the energetically preferred solution for large enough $\mu$. As found in \cite{rnt}, there is a critical value of $\mu = \mu_c < \mu_0$ below which the scalar field condenses. For $d=3$ and $m^2=-15/4$,  the value of $\mu_c \sim 1.89/q$, so that for large enough $q$ this is smaller than $\mu_0$ when the background is AdS soliton.

The point $\mu = \mu_c$ is a critical point with diverging correlation length. An important property of this point is that there is a scalar zero mode, i.e. a solution of the linearized equations of motion for the scalar which satisfies the vanishing source boundary condition at the AdS boundary, and in addition regular in the interior. For specific values of $m^2$, the existence of the zero mode was proved in \cite{bddn}. For generic $m^2$ in this mass range, this has been found numerically. This zero mode will play a key role in the following.

The existence of this zero mode allows a calculation of the equilibrium critical exponents. In the following we will be interested in a quantum quench driven by a time dependent boundary value of the scalar field $J(t)$ with $\mu$ tuned to be exactly equal to $\mu_c$. We therefore need to know the equilibrium exponents at $\mu = \mu_c$. For static configurations the equations of motion (\ref{eom_phi_R_ads}) - (\ref{eom_at_ads}) become
\begin{align}
D \Phi_R + \frac{1}{r^4h}  (\phi^2-\mu^2) \Phi_R 
 = 0,
\label{eom_phi_R_ads_static}
 \\
D \Phi_I 
 +  \frac{1}{r^4h} (\phi^2-\mu^2) \Phi_I 
 = 0,
 \\
0= 2 r^2 \left( \Phi_I \Phi_R' - \Phi_I' \Phi_R   \right),
\label{eom_ar_ads_static} \\
\frac{1}{r^d} \partial_r \left( r^{d} h \partial_r  \phi \right)
=\frac{2}{r^2}  \phi |\Phi|^2  .
\label{eom_at_ads_static} 
\end{align}
where the operator $D$ is given by
\begin{align}
D \equiv  \partial_r^2+ \left(  \frac{d+2}{r}+ \frac{h'}{h} \right) \partial_r  
 - \frac{1}{r^2h} \left(m^2  - \frac{\mu^2}{r^2}\right) .
\label{op_D}
\end{align}
Consider a time independent boundary condition given by $J$ with $\mu =\mu_c$. By a $r$ independent gauge transformation we can choose the static field $\Phi$ to be real - this satisfies the constraint equation (\ref{eom_ar_ads_static}) for time independent configurations. To calculate the response to such a static source in the critical region $|J| \ll 1$ let us expand the fields as
\begin{align}
\Phi_R = J f(r) + J^{\gamma} \chi (r), \quad \phi=\mu + J^{\beta} \alpha(r),
\label{2-1}
\end{align} 
where $f(r)= r^{-\Delta_-}(1 + \cdots ),  \chi(r) \sim r^{-\Delta_+}$  and 
$\alpha (r) \sim r^{-(d-1)}$
 for large $r$. The functions $\chi (r)$ and $\alpha (r)$ have an expansion in powers of $J$ and we will seek a solution which starts at $O(J^0)$. We will assume that $\gamma \le 1$. The self-consistency of this assumption will be verified below. Substituting (\ref{2-1}) in equations (\ref{eom_phi_R_ads_static}) and (\ref{eom_at_ads_static}) we get
\begin{align}
J D(f) + J^{\gamma} D(\chi)
+ \frac{1}{r^4h}  ( 2 \mu J^{\beta} \alpha + J^{2\beta}\alpha^2 )(Jf + J^\gamma \chi)
 = 0,
\label{eom_phi_R_ads_static_2}
 \\
J^{\beta} \frac{1}{r^d} \partial_r \left( r^{d} h \partial_r  \alpha \right)
=\frac{2}{r^2}(Jf + J^{\gamma} \chi)^2( \mu + J^\beta \alpha).
\label{eom_at_ads_static_2} 
\end{align}
For small $J$, it is straightforward to see that when $D\chi \neq 0$ there are $O(J^0)$ solutions to these equations only when  $\gamma =1$ and $ \beta = 2$. However for $\mu = \mu_c$ there is a zero mode, i.e. a solution to $D\chi =0$. In this case the only possibility is $\gamma < 1$ and the equations then determine 
\ben
\beta = 2/3,~~~~~~~~~~\gamma = 1/3.
\een
Since the term $J^\beta a$ is the expectation value of the charge density and the term $J^\gamma \chi$ is the expectation value of the order parameter in the boundary theory, we get
\begin{align}
\langle \cO \rangle \sim J^{1/3}, \quad \langle \rho \rangle \sim J^{2/3}.
\end{align}
We therefore get the mean field exponents appropriate to a $\Phi^4$ landau-Ginsburg theory. 

\subsection{The Adiabatic Expansion and its Breakdown}
\label{adiabatic}

We want to study the response of the system at $\mu = \mu_c$ in the presence of a time dependent boundary condition $J(t)$ (see equation (\ref{asymptotics-phi}) for the bulk scalar). This is dual to a time dependent source for the order parameter in the boundary theory. The function $J(t)$ is chosen to asymptote to a constant value at early and late times, and slowly varying compared to the initial gap, which has been set to unity by the choice $r_0=1$. $J(t)$ crosses zero at some intermediate time. An example is
\ben
J(t) = J_0 \tanh (vt), ~~~~~~~~v \ll 1.
\label{2-3-1}
\een
Let us consider starting the system at an early time where $J(t) \sim -J_0$. For early enough times, the time evolution will be adiabatic. However, as $t \rightarrow 0$, the system approaches a critical point with a vanishing (instantaneous) gap, leading to a breakdown of adiabaticity. In this subsection we will determine the time scale at which this happens.

An adiabatic solution to the equations of motion (\ref{eom_phi_R_ads}) - (\ref{eom_at_ads}) has the form
\begin{align}
\Phi_R &= \Phi_R^{(0)}(r,t)+ \epsilon \Phi_R^{(1)}(r,t)+ \epsilon^2 \Phi_R^{(2)}(r,t) + \cdots ,\nonumber \\
\Phi_I &= \Phi_I^{(0)}(r,t) +\epsilon \Phi_I^{(1)}(r,t)+\epsilon^2 \Phi_I^{(2)}(r,t) + \cdots ,\nonumber \\
\phi &= \phi^{(0)}(r,t) + \epsilon \alpha^{(1)}(r,t) + \epsilon^2 \alpha^{(2)}(r,t) \cdots ,
\label{2-3-2}
\end{align}
where $\epsilon$ is the adiabaticity parameter which counts the number of time derivatives, and the functions $\Phi_R^{(0)}, \Phi_I^{(0)}, \phi^{(0)}$ are the static solutions discussed in the previous section, but with the constant source $J$ replaced by $J(t)$,
\begin{align}
\Phi_R^{(0)}(t,r) & =  J(t) f_0(r) + J(t)^{1/3} \chi^{(0)}(r),~~~~~~ \Phi_I^{(0)} = 0, \\
\phi^{(0)}(t,r) & =   \mu + J(t)^{2/3} \alpha^{(0)}(r).
\label{2-3-3}
\end{align}
The functions appearing above have the following asymptotic behavior at $r \rightarrow \infty$
\ben
f_0(r) \sim r^{-\Delta_-},~~~\chi^{(0)}(r) \sim r^{-\Delta_+},~~~
\alpha ^{(0)}(r) \sim r^{1-d}.
\label{2-3-4}
\een
We now substitute (\ref{2-3-2}) into the equations of motion (\ref{eom_phi_R_ads}) - (\ref{eom_at_ads}), replace $\partial_t \rightarrow \epsilon \partial_t$ and equate terms with the same power of $\epsilon$. The $O(\epsilon)$ equations become
 \begin{align}
D(\Phi_R^{(1)})
+  \frac{1}{r^4h} \left(2 J^{2/3} \mu \alpha^{(0)} \Phi_R^{(1)}  + 2 J^{1/3} \mu \alpha^{(1)} \chi^{(0)} \right) = 0,
\label{eom_phi_R_ads_1}
 \\
D(\Phi_I^{(1)})
-  \frac{1}{r^4h} \left(   -2 J^{2/3} \mu \alpha^{(0)} \Phi_I^{(1)} \right) 
 =    - \frac{2}{3r^4h} \mu J^{-2/3} \dot J \chi^{(0)},  
\label{eom_phi_I_ads_1}
 \\
\frac23 J^{-1/3} \dot J \partial_r \alpha^{(0)}= 2 r^2 \left( \Phi_I^{(1)} \Phi_R^{'(0)} - \Phi_I^{'(1)} \Phi_R^{(0)}   \right)
\label{eom_ar_ads_1} ,\\
\frac{1}{r^d} \partial_r \left( r^{d} h \partial_r  \alpha^{(1)} \right)
=\frac{2}{r^2} \left( 2 \mu  J^{1/3} \chi^{(0)} \Phi_R^{(1)} + \alpha^{(1)} J^{2/3}  \left( \chi^{(0)} \right)^2  \right).
\label{eom_at_ads_1} 
\end{align}
In the above equations we have retained the leading order terms for small $J$ since this is the regime where we expect adiabaticity to break down.
The equations (\ref{eom_phi_R_ads_1}) and (\ref{eom_at_ads_1}) are {\em homogeneous} coupled equations for $\Phi_R^{(1)}$ and $\alpha^{(1)}$. With the specified boundary conditions their solutions are trivial
\ben
\Phi_R^{(1)} = \alpha^{(1)} = 0,
\label{2-3-5}
\een
at the lowest order of the small $J$ expansion.

On the other hand, the equation (\ref{eom_phi_I_ads_1}) has a source which is the time derivative of the zeroth order solution. Since the background has $\mu = \mu_c$, the operator $D$ has a zero mode. If $J$ is vanishing, the solution to this equation is therefore divergent, signifying a breakdown of adiabaticity. For small $J$, we can use perturbation theory to estimate $\Phi_I^{(1)}$. To do this it is convenient to decompose the field $\Phi_I^{(1)}(r,t)$ in terms of the orthonormal eigenfunctions of the operator $D$:
\ben
\Phi_I^{(1)}(r,t) = \sum_n \Phi_{I,n}^{(1)}(t) \varphi_n (r),
\quad D\varphi_n(r) = \lambda_n \varphi_n (r),
\quad (\lambda_0=0, \quad \lambda_n >0 \quad (n \ge  1) )
\label{2-3-5-a}
\een
The equation (\ref{eom_phi_I_ads_1}) can be then expressed in the form
\ben
\lambda_n \Phi_{I,n}^{(1)}(t) + J^{2/3} {\tilde{\cA}}^m_n \Phi_{I,m}^{(1)}(t) = J^{-2/3} {\dot J} {\tilde{\cB}}_n
\label{2-3-5-b}
\een
where we have defined 
\ben
{\tilde{\cA}}^m_n = 2\mu \int dr \varphi^\star_n(r)  \frac{1}{r^4h} \alpha^{(0)} \varphi_m (r)~~~~~~~{\tilde{\cB}}_n =-\mu \int dr \frac{2}{3r^4h} \chi^{(0)}(r) 
\label{2-3-5-c}
\een
It is clear from (\ref{2-3-5-b}) that for small $J(t)$ while the contribution from the nonzero modes begin with $J^{-2/3} {\dot J}$ the zero mode contribution is proportional to $J^{-4/3} {\dot J}$. Thus the dominant adiabatic correction is given by 
\ben
\Phi_I^{(1)} \sim J^{-4/3} {\dot J}.
\label{2-3-6}
\een
The adiabatic expansion then breaks down when $\Phi_I^{(1)}$ is of the same order as $\Phi_R^{(0)}$, i.e. when
\ben
{\dot J} \sim J^{5/3}.
\een
For a generic protocol with $J(t) \sim vt$ for small $t$, this means that the Kibble-Zurek time is 
\ben
t_\star \sim v^{-2/5},
\label{2-3-8}
\een
while the order parameter at this time is
\ben
\langle \cO \rangle \sim v^{1/5}.
\label{2-3-9}
\een

We need to make sure that the equation (\ref{eom_ar_ads_1}) is consistent with the small $J$ behavior above. Naively the $J$ dependence of $\Phi_I^{(1)}$ is not consistent with (\ref{eom_ar_ads_1}). However, as mentioned earlier this equation is a constraint equation and it is sufficient to check this at large $r$, i.e. check the equation (\ref{constrainteqn}). Using the results ${\rm Im}J = 0$ and 
\ben
\rho \sim J^{2/3},~~~{\rm Im}\chi \sim J^{-4/3}{\dot J},~~~
{\rm Re}J \sim J,
\een
it is clear that both sides of this equation behave as $J^{-1/3}{\dot J}$.
This agreement reflects the fact that there are cancellations in the right hand side 
of (\ref{eom_ar_ads_1}) as is explicit in the derivation of (\ref{constrainteqn}).

We therefore see that to lowest order in the adiabatic expansion, the gauge field does not receive any correction. To investigate any breakdown of adiabaticity in the gauge sector, we need to proceed to the next order in the adiabatic expansion. To $O(\epsilon^2)$ the equation of motion (\ref{eom_at_ads}) leads to
\begin{align}
\frac{1}{r^d}\partial_r(r^d h \partial_r \alpha^{(2)})  = & \frac{2}{r^2}\left[\Phi_I^{(1)}\partial_t \Phi_R^{(0)} - \Phi_R^{(0)}\partial_t \Phi_I^{(0)} + \alpha^{(0)}(\Phi_I^{(1)})^2 \right] \nonumber \\
& + \frac{2}{r^2} \left[ \alpha^{(2)}(\Phi_R^{(0)})^2 + 2 \alpha^{(0)}\Phi_R^{(0)}\Phi_R^{(2)}\right],
\label{2-3-7}
\end{align}
where we have used $\Phi_R^{(1)}=\Phi_I^{(0)}=0$. For small $J$ we have $\Phi_I^{(1)} \sim J^{-4/3}{\dot J}$ and $\Phi_R^{(0)} \sim J^{1/3}$. The equation becomes
\begin{align}
\frac{1}{r^d}\partial_r(r^d h \partial_r \alpha^{(2)}) = &K_1(r) J^{-1}{\ddot J}+K_2(r) J^{-2} {\dot J}^2+ \mu K_3(r) J^{-8/3} {\dot J}^2 \nonumber \\
& +J^{2/3}L_1(r) \alpha^{(2)}+ \mu J^{1/3} L_2(r) \Phi_R^{(2)},
\end{align}
where $K_i(r), L_i(r)$ are functions of $r$. Unlike the case of the scalar, the operator on the left hand side does not have a zero mode. Therefore we can ignore the terms which have positive powers of $J$. The equation then becomes a linear differential equation for $\a^{(2)}$ with a source which arises from the first order corrections. These sources clearly diverge in the $J \rightarrow 0$ limit, so that there are large corrections to the gauge field as $J$ approaches zero, and $\a^{(2)}$ scales as
\ben
\alpha^{(2)} \sim J^{-8/3} \dot J^2.
\een
Thus $\alpha^{(2)} \sim J^{2/3} \sim v^{2/5} (\sim J^{2/3} \alpha^{(0)})$ at the Kibble-Zurek time (\ref{2-3-8}) and the adiabaticity of the charge density $\rho$ is broken.
This means that the breakdown of adiabaticity in the scalar sector feeds into the gauge sector.

\subsection{Scaling in the Critical Region}
\label{sec-scaling}

Once adiabaticity breaks down, there is no Taylor series expansion in
$v$ as in (\ref{2-3-2}). We now show that there is now a different small-$v$ expansion, in
powers of $v^{2/5}$. To see this, it is convenient to rescale the time
\ben 
t \rightarrow \eta = v^{2/5} t,
\label{2-4-1}
\een
and separate out the source part of the fields as follows
\begin{align}
\Phi_R  & =  J(\eta v^{-2/5}) r^{-\Delta_-} + v^{1/5} \chi_R (\eta,r), \\   
\Phi_I  & =  v^{1/5} \chi_I(\eta,r) ,\\
\phi  & =  \mu + v^{2/5} \alpha(\eta,r).
\label{2-4-2}
\end{align}
For large $r$ both $\chi_R, \chi_I \sim r^{-\Delta_+}$ and $\alpha \rightarrow r^{-{d-1}}$.

Near $t =0$ we replace 
\ben
J(t) \rightarrow J_0 (vt) = J_0 v^{3/5}\eta.
\een
The equations of motion (\ref{eom_phi_R_ads})-(\ref{eom_at_ads}) can
be now expanded in powers of $v$ and become
\begin{align}
v^{1/5}D(\chi_R)+v^{3/5} \left[ -  \frac{1}{r^4h} \left(  2 \mu
  \partial_\eta \chi_I - 
2\mu \alpha \chi_R \right) + J_0 \eta D(r^{-\Delta_-}) \right] + O(v) = 0,
\label{eom_phi_R_scale}
 \\
 v^{1/5} D (\chi_I)+
v^{3/5} \left[
-  \frac{1}{r^4h} \left( - 2 \mu  \partial_\eta \chi_R -2 \mu \alpha \chi_I \right)  \right] + O(v)
 = 0,
\label{eom_phi_I_scale}
 \\
v^{4/5} \dot \alpha '= 2 r^2 J_0 \eta v^{4/5} \left[ \chi_I
  \partial_r(r^{-\Delta_-}) - (\partial_r \chi_I )r^{-\Delta}
  \right] + 2r^2 v^{2/5} (\chi_I\partial_r \chi_R - \chi_R \partial_r \chi_I),
\label{eom_ar_scale} \\
v^{2/5} \left[ \frac{1}{r^d} \partial_r \left( r^{d} h \partial_r  \alpha \right)-\frac{2}{r^2}   \mu  \left(\chi_R^2+\chi_I^2  \right) \right] + O(v^{4/5})
=0.
\label{eom_at_scale} 
\end{align}
Here $\cdot$ denotes $\eta$ derivative.
As mentioned earlier, the equation (\ref{eom_ar_scale}) is a constraint equation which needs to be imposed at some $r$, the other equations then guarantee that it holds for all $r$. It is straightforward to check that the $v^{2/5}$ term in (\ref{eom_ar_scale}) vanishes for large $r$.

To solve these equations, we first consider equation (\ref{eom_at_scale}).
 Since the operator
$D_1 \equiv \frac{1}{r^d} \partial_r \left( r^{d} h \partial_r \right)$ does not have a zero mode, (\ref{eom_at_scale}) can be solved by obtaining the relevant Green's function
\begin{align}
\alpha = D_1^{-1}\left[ \frac{2}{r^2}   \mu  \left(\chi_R^2+\chi_I^2  \right)  \right].
\label{2-4-8}
\end{align}

We next consider the equations of motion for scalars (\ref{eom_phi_R_scale}) and (\ref{eom_phi_I_scale}). 
Since we are working at $\mu = \mu_c$, the operator $D$ has a zero mode. It is clear that the zero mode dominates the scalar dynamics for small $v$.
More precisely, consider expanding the fields in the basis formed by the eigenvectors of $D$ which we employed in (\ref{2-3-5-a}),
\begin{align}
 D\varphi_n(r) = \lambda_n \varphi_n(r), \quad (\lambda_0=0, \quad \lambda_n >0 \quad (n \ge  1) ).
\end{align}
Hence,
\begin{align}
& \chi_{R} (\eta,r) = \sum_{n} \chi_{R,n} (\eta) \varphi_n (r),  \\
& \chi_{I} (\eta,r) = \sum_{n} \chi_{I,n} (\eta) \varphi_n (r) , \\
& \alpha (\eta,r) =  \sum_{n} \alpha_n (\eta) \varphi_n (r).
\end{align}
Then e.g. the equations (\ref{eom_phi_R_scale}) and (\ref{eom_phi_I_scale}) may be written as infinite sets of ordinary differential equations
\begin{align}
& \lambda_p \chi_{R,p} = v^{2/5} \left[ 2\mu \cA_p^n(\partial_\eta \chi_{I,n}) - 2\mu \cB_p^{mn} \alpha_m \chi_{R,n} - J_0 \eta \cJ_p \right] + O(v^{4/5}), \\
& \lambda_p \chi_{I,p} = v^{2/5} \left[- 2\mu \cA_p^n(\partial_\eta \chi_{R,n}) - 2\mu \cB_p^{mn} \alpha_m \chi_{I,n} \right] + O(v^{4/5}),
\end{align}
where
\begin{align}
&\cA_p^n = \int [dr]\frac{1}{r^4h}\varphi^\star_p(r)\varphi_n(r) ,\\
&\cB_p^{mn} =  \int [dr]\frac{1}{r^4h}\varphi^\star_p(r)\varphi_m(r)\varphi_n(r) ,\\
&\cJ_p =  \int [dr] (Dr^{-\Delta_-})\varphi^\star_p(r),
\end{align}
and $[dr]$ denotes the measure with which the eigenfunctions are orthonormal.

Clearly these equations have solutions which have an expansion in powers of $v^{2/5}$. 
The zero mode $\chi_{R,0}$ has a $O(1)$ contribution, while the dominant contribution to the non-zero modes is $O(v^{2/5})$. The small-$v$ dynamics is therefore given the following set of equations
\begin{align}
 -2\mu \cA_0^0(\partial_\eta \chi_{I,0}) + 2\mu \cB_0^{0m} \alpha^0_m \chi_{R,0} + J_0 \eta \cJ_0 & = 0 ,\\
2\mu \cA_0^0(\partial_\eta \chi_{R,0}) + 2\mu \cB_0^{0m} \alpha^0_m\chi_{I,0} & = 0,
\end{align}
while $\{ \alpha^0_m \}$ are determined by the equation (\ref{2-4-8}) with $\chi_R \rightarrow \chi_{R,0}$ and  $\chi_I \rightarrow \chi_{I,0}$.  

Going back to the original variables this means that the leading order solutions for the {\em normalizable} parts of the bulk fields have the scaling forms
\begin{align}
\Phi_{R,I}(t,v) & = v^{1/5}\Phi_{R,I} (tv^{2/5},1), \\
\phi(t,v) & = v^{2/5} \phi(tv^{2/5},1).
\end{align}
This implies that the expectation values of the order parameter $\langle \cO \rangle$ and the charge density $\langle \rho \rangle$ in the boundary theory obey the Kibble-Zurek type scaling relations
\begin{align}
\langle \cO (t,v)\rangle & = v^{1/5} F(tv^{2/5}), \\
\langle \rho(t,v)\rangle & = v^{2/5} G(tv^{2/5}).
\end{align}

\section{Gravity Backreaction}\label{sec4}
\setcounter{equation}{0}

In this section we consider the backreaction of the metric for the case $d=3$. A similar discussion will hold for other dimensionalities. The set-up is the same as in the previous section. We consider the system with the chemical potential tuned exactly to $\mu = \mu_c$ and then turn on a source for the order parameter which is a function of time only.  The full equations of motion are now given by
\begin{align}
\nabla^2 \Phi =& m^2 \Phi,
\label{eom-scalar}
 \\
\frac{1}{\sqrt{-g}} \partial_\mu \sqrt{-g} g^{\mu\nu} g^{\rho \sigma} F_{\nu \rho} =& J^{\sigma} ,\\
R_{\mu\nu}-\frac{1}{2}g_{\mu\nu}R-6 g_{\mu\nu}
=& \frac{1}{2} F_{\mu\lambda} {F_{\nu}}^{\lambda} +
\frac12 \left( \nabla_\mu \Phi \nabla_\nu \Phi^*+\nabla_\nu \Phi \nabla_\mu \Phi^* \right) \nonumber \\
&-\frac{g_{\mu\nu}}{2}\left(
\frac{1}{4} F_{\rho\sigma} F^{\rho\sigma} + m^2 |\Phi|^2 + |D \Phi|^2
\right).
\label{eom-gr}
\end{align}
where we have put $\kappa^2=1/2$.
The symmetries of the system allow gauge choices leading to the following forms of the fields \cite{horo}
\begin{align}
ds^2 & =r^2\left(e^{A(r,t)}B(r,t)d\theta^2+dx^2+dy^2-e^{C(r,t)}dt^2\right)+\frac{dr^2}{r^2B(r,t)},
\label{metric-ansatz} \\
A_t & =\phi(r,t),\qquad \Phi=\Phi(r,t).
\label{Apsiansatz}
\end{align}
The complete equations of motion are given in the Appendix.

\subsection{Static Solutions and Scaling}

When the fields are independent of time, the equations of motion simplify \cite{horo}.  The coupled Maxwell-scalar equations become 
\begin{equation}
\Phi''+\left(\frac{5}{r}+\frac{A'}{2}+\frac{B'}{B}+\frac{C'}{2}\right)\Phi'+\frac{1}{r^2B}\left(\frac{e^{-C}\phi^2}{r^2}-m^2\right)\Phi=0\;,
\label{eom_phi_gr}
\end{equation}
\begin{equation}
\phi''+\left(\frac{3}{r}+\frac{A'}{2}+\frac{B'}{B}-\frac{C'}{2}\right)\phi'-\frac{2\Phi^2\phi}{r^2B}=0\;.
\label{eom_gauge_gr}
\end{equation}
The nontrivial components of the Einstein equations (\ref{eom-gr}) are the $tt$, $rr$, $\theta\theta$ and $xx$ components.
However one of them is the constraint equation and we have only three dynamical equations.
Following \cite{horo}, we take linear combinations of these equations.
From $g^{\theta\theta}(G_{\theta\theta}-T_{\theta\theta})-g^{rr}(G_{rr}-T_{rr})=0$ (where $G_{AB}$ is the Einstein tensor and $T_{AB}$ is the bulk energy momentum tensor),  we obtain 
\begin{equation}
A'=\frac{2r^2C''+r^2C'^2+4rC'+4r^2\Phi'^2-2e^{-C}\phi'^2}{r(6+rC')}\;,
\label{eom_A_gr}
\end{equation}
and, from $g^{xx}(G_{xx}-T_{xx})-g^{tt}(G_{tt}-T_{tt})=0$, we obtain
\begin{equation}
C''+\frac12 C'^2+\left(\frac{5}{r}+\frac{A'}{2}+\frac{B'}{B}\right)C'-\left(\phi'^2+\frac{2\phi^2\Phi^2}{r^2B}\right)\frac{e^{-C}}{r^2}=0
\label{eom_C_gr}.
\end{equation}
Finally from $g^{xx}(G_{xx}-T_{xx})-g^{tt}(G_{tt}-T_{tt})-g^{\theta\theta}(G_{\theta\theta}-T_{\theta\theta})=0$, we obtain
\begin{align}
B'\left(\frac{3}{r}-\frac{C'}{2}\right)+B\left(\Phi'^2-\frac12 A'C'+\frac{e^{-C}\phi'^2}{2r^2}+\frac{12}{r^2}\right) \nonumber \\
+\frac{1}{r^2}\left(\frac{e^{-C}\phi^2\Phi^2}{r^2}+m^2\Phi^2-12\right)=0\;.
\label{eom_B_gr}
\end{align}

The static $AdS$ soliton solution is given by
\begin{align}
\Phi=0, \quad
\phi=\mu, \quad A=0,\quad B=h(r)\equiv 1- \left(\frac{r_0}{r}\right)^4, \quad C=0
\label{3-2-1}.
\end{align}
There is a critical value $\mu = \mu_c$ such that for $\mu > \mu_c$ this is not the favored solution - rather the solution is a hairy soliton which has been found in \cite{horo}.
This is a solution with a vanishing source, i.e. the fields do not have a non-normalizable piece. We are, however, interested in solutions with a source $J$. We will work exactly at $\mu = \mu_c$ so that for $J \ll 1$ the departure from the solution (\ref{3-2-1}) is small. The fields can be then expanded as
\begin{align}
\Phi (r) = J f(r)+ J^{\gamma} \chi (r),\\
\phi (r) = \mu + J^{\beta} \alpha (r) ,\\
A(r) =J^{\delta} a(r),\\
B(r) =h (r) +J^{\epsilon} b (r) ,\\
C(r) =J^{\eta} c (r).
\end{align}
The strategy is to now look at the static equations at the lowest nontrivial order of $J$ and look for solutions for $\chi,\alpha,a,b,c$ which start at $O(J^0)$ .
The leading terms in  (\ref{eom_gauge_gr}) yield
\begin{align}
J^{\beta}\alpha''+\left(\frac{3}{r}+\frac{h'}{h}\right)J^{\beta} \a'-\frac{2 \mu  J^{2\gamma}\chi^2}{r^2 h}=0\;.
\end{align}
Thus there is a $O(J^0)$ solution for $\alpha,  \chi$ if $\beta = 2\gamma$. Similarly
(\ref{eom_C_gr}) leads to 
\begin{equation}
J^{\eta}c''+\left(\frac{5}{r}+\frac{h'}{h}\right)J^{\eta}c'-\left(\frac{2 \mu^2 J^{2\gamma}\chi^2}{r^2h}\right)\frac{1}{r^2}=0,
\end{equation}
which implies $\eta = 2 \gamma$.
To leading order the equation (\ref{eom_A_gr}) gives
\begin{equation}
J^{\delta}a'=\frac{ 2r^2 J^{2\gamma} c''+4r J^{2\gamma} c'+4r^2 J^{2\gamma} \chi'^2}{6r}\;.
\end{equation}
This leads to  $\delta = 2 \gamma$.
The component of the Einstein equations (\ref{eom_B_gr}) becomes
\begin{align}
\frac{3}{r}J^{\epsilon}b'+h'\left(-\frac{J^{2\gamma}c'}{2}\right)+h\left(J^{2\gamma}\chi'^2\right) 
+\frac{12}{r^2} J^{\epsilon} b 
+\frac{1}{r^2}\left(\frac{\mu^2}{r^2}+m^2 
\right)J^{2\gamma}\chi^2=0\;,
\end{align}
so that we obtain $\epsilon = 2 \gamma$.
Finally the scalar equation of motion (\ref{eom_phi_gr}) becomes
\begin{align}
Jf''+J^{\gamma}\chi''+\left(\frac{5}{r}+\frac{h'}{h}\right)(Jf'+J^{\gamma}\chi)
+J^{2\gamma}\left(\frac{a'}{2}+\frac{b'}{h}-\frac{h' b}{h^2}+\frac{c'}{2}\right)J^{\gamma}\chi' \nonumber \\
+\frac{1}{r^2h}\left(\frac{\mu^2}{r^2}-m^2\right)(Jf'+J^{\gamma}\chi)
+\frac{J^{2\gamma}}{r^2h}\left(\frac{ 1}{r^2}(2\mu \alpha -\mu^2c)\right)J^{\gamma}\chi \nonumber \\
-\frac{J^{2\gamma} b}{r^2h^2}\left(\frac{\mu^2}{r^2} -m^2\right)J^{\gamma}\chi
=0.
\label{3-2-2}
\end{align}
Using the definition of the operator $D$
\begin{align}
D(f) \equiv f''+\left(\frac{5}{r}+\frac{h'}{h}\right)f'
+\frac{1}{r^2h}\left(\frac{\mu^2}{r^2}-m^2\right)f',
\label{D-gravity}
\end{align}
 this becomes
\begin{align}
JD(f)+J^{\gamma}D(\chi)
+J^{3\gamma}\left(\frac{a'}{2}+\frac{b'}{h}-\frac{h' b}{h^2}+\frac{c'}{2}\right)\chi' \nonumber \\
+\frac{J^{3\gamma}}{r^2h}\left(\frac{ 1}{r^2}(2\mu \alpha -\mu^2c)\right)\chi 
-\frac{J^{3\gamma} b}{r^2h^2}\left(\frac{\mu^2}{r^2} -m^2\right)\chi
=0.
\label{3-2-4}
\end{align}
Since we are exactly at $\mu =\mu_c$, the operator $D$ has a zero mode. For this mode the first two terms in (\ref{3-2-4}) vanish and comparing the last three terms one immediately obtains
$\gamma =1/3$.
Thus the results are summarized as
\begin{align}
\Phi (r) =& J f (r)  + J^{1/3} \chi (r), \\
\phi (r) =& \mu + J^{2/3} \alpha (r) ,\\
A (r) =&J^{2/3} a (r) ,\\
B (r) =&h (r) +J^{2/3} b (r), \\
C (r) =&J^{2/3} c (r).
\end{align}
It can be checked that for small $J$ the usual relation between the subleading pieces of the gauge field and the metric with the boundary theory current and energy momentum tensor expectation values is unchanged. Therefore the critical behavior of the order parameter $\langle O \rangle$, the charge density and the energy momentum tensor are
 \begin{align}
\langle O \rangle \sim J^{1/3},~~~~~~~~~
\langle \rho \rangle \sim J^{2/3},~~~~~~~~~\langle T_{\mu\nu} \rangle \sim J^{2/3}.
\end{align}

\subsection{Adiabaticity Breakdown}

We now follow the treatment of section (\ref{adiabatic}) to investigate the manner in which adiabaticity breaks for a time dependent source $J(t)$ as we approach the critical point at $J=0$. The derivative expansions for the various fields are
\begin{align}
\Phi_R &= \Phi_R^{(0)}+ \epsilon \Phi_{R}^{(1)}+ \epsilon^2 \Phi_{R}^{(2)} + \cdots \nonumber ,\\
\Phi_I &= \Phi_I^{(0)}+ \epsilon \Phi_{I}^{(1)}+ \epsilon^2 \Phi_{I}^{(2)} + \cdots \nonumber ,\\
A_t &= \phi^{(0)}+ \epsilon \alpha^{(1)}+\epsilon^2\alpha^{(2)} +  \cdots \nonumber ,\\
A &= A^{(0)}+ \epsilon a^{(1)}+\epsilon^2 a^{(2)} + \cdots, \nonumber \\
B &= B^{(0)}+ \epsilon b^{(1)}+ \epsilon^2 b^{(2)} + \cdots ,\nonumber \\
C &= C^{(0)}+ \epsilon c^{(1)}+ \epsilon^2 c^{(2)} +\cdots,
\label{3-3-8}
\end{align}
where $\epsilon$ is the adiabaticity parameter and the lowest order solutions are obtained from the static solutions by replacing the constant source $J$ by the time dependent source $J(t)$,
\begin{align}
\Phi_R^{(0)}(r,t)  & = J(t) f_0 (r) + J(t)^{1/3} \chi_{R}^{(0)} (r),\\
\Phi_I^{(0)}(r,t) & = 0, \\
\phi^{(0)} (r,t) & = \mu + J(t)^{2/3} \alpha^{(0)} (r), \\
A^{(0)}(r,t) & = J(t)^{2/3} a^{(0)} (r), \\
B^{(0)}(r,t) & = h (r) + J(t)^{2/3} b^{(0)}(r), \\
C^{(0)}(r,t) & = J(t)^{2/3} c^{(0)} (r) .
\label{3-3-1}
\end{align}
The equations which determine the adiabatic corrections are obtained from the full equations of motion in the Appendix, replacing $\partial_t \rightarrow \epsilon \partial_t$, and equating terms of a given order in $\epsilon$. To $O(\epsilon)$, the real part of the scalar equation of motion, the Maxwell equation, and the combinations of $(rr),(tt),(xx)$ and $(\theta\theta)$ components of the Einstein equations do not contain any time derivatives and form a set of {\em homogeneous} coupled differential equations for $\Phi_R^{(1)}, \alpha^{(1)},a^{(1)},b^{(1)}$ and $c^{(1)}$. 
For example the real part of the scalar field equation yields
\begin{align}
&\Phi_R''+\left(\frac{5}{r}+\frac{A'}{2}+\frac{B'}{B}+\frac{C'}{2}\right)\Phi_R'+\frac{1}{r^2B}\left(\frac{e^{-C}\phi^2}{r^2}-m^2\right)\Phi_R \nonumber \\
&-\frac{e^{-C}}{r^4 B}\left[
\ddot \Phi_R + \frac{\dot A- \dot C}{2} \dot \Phi_R
+2\phi \dot \Phi_I + \dot \phi \Phi_I+ \frac{\phi}{2}(\dot A- \dot C)\Phi_I
\right]
=0\;.
\label{3-3-6}
\end{align}
Since $\Phi_I^{(0)}=0$, all the terms which contain time derivatives in (\ref{3-3-6}) are at least $O(\epsilon^2)$. 
We can now expand the fields which appear in the first two lines of (\ref{3-3-6}) in powers of $\epsilon$. To $O(\epsilon)$ this yields
\begin{align}
(\Phi_{R}^{(1)})^{\prime\prime} + & \left(\frac{5}{r}+\frac{(A^{(0)})^\prime}{2} + \frac{(B^{(0)})^\prime}{B^{(0)}} + \frac{(C^{(0)})^\prime}{2} \right) 
(\Phi_{R}^{(1)})^{\prime} \nonumber \\
& + \left(\frac{(a^{(1)})^\prime}{2} +\frac{(b^{(1)})^\prime}{B^{(0)}}-
\frac{(B^{(0)})^\prime b^{(1)}}{(B^{(0)})^2} + \frac{(c^{(1)})^\prime}{2} \right) (\Phi_{R}^{(0)})^{\prime}\nonumber \\
& +  \frac{1}{r^2 B^{(0)}} \left( \frac{e^{-C^{(0)}}}{r^2} \left(2\phi^{(0)}\phi^{(1)} - c^{(1)}(\phi^{(0)})^2 -\frac{b^{(1)}}{B^{(0)}} (\phi^{(0)})^2 \right) + m^2\frac{b^{(1)}}{B^{(0)}}\right)\Phi_R^{(0)}\nonumber \\
& + \frac{1}{r^2 B^{(0)}} \left( \frac{e^{-C^{(0)}}(\phi^{(0)})^2}{r^2}-m^2 \right)\Phi_R^{(1)} = 0.
\label{3-3-2}
\end{align}
Similarly the Maxwell's equation (\ref{eom_gauge_gr_full}) and the diagonal components of the Einstein equations (\ref{eom_A_gr_full}) - (\ref{eom_B_gr_full}) lead to homogeneous linear differential equations for
$\Phi_R^{(1)}, \alpha^{(1)},a^{(1)},b^{(1)}$ and $c^{(1)}$.
Since these corrections  have to satisfy regularity conditions in the interior as well as normalizable boundary conditions at the boundary, and the equations which govern them do not involve any inhomogeneous term, the solutions are trivial, i.e.
\ben
\Phi_R^{(1)} = \alpha^{(1)} = a^{(1)} = b^{(1)} = c^{(1)} = 0.
\een

The equations which involve time derivatives of the zeroth order fields are those which follow from the imaginary part of the scalar field equation,  the $(rt)$ component of the Einstein equations and the $r$ component of the Maxwell equations which is the constraint equation corresponding to (\ref{eom_ar_ads_1}). The imaginary part of the scalar equation reads
\begin{align}
&\Phi_I''+\left(\frac{5}{r}+\frac{A'}{2}+\frac{B'}{B}+\frac{C'}{2}\right)\Phi_I'+\frac{1}{r^2B}\left(\frac{e^{-C}\phi^2}{r^2}-m^2\right)\Phi_I \nonumber \\
&-\frac{e^{-C}}{r^4 B}\left[
\ddot \Phi_I + \frac{\dot A- \dot C}{2} \dot \Phi_I
-2\phi \dot \Phi_R - \dot \phi \Phi_R- \frac{\dot A- \dot C}{2}\phi\Phi_R
\right]
=0\;.\label{3-3-7}
\end{align}
We now substitute the adiabatic expansions (\ref{3-3-8}) with the leading order fields given by (\ref{3-3-1}) and retain the lowest order terms in a small $J$ expansion. This may be written in terms of the operator $D$ introduced in (\ref{D-gravity}),
\begin{align}
D(\Phi_I^{(1)})
& +J^{2/3}\left(\frac{{a^{(0)}}'}{2}+\frac{{b^{(0)}}'}{h}-\frac{h' b^{(0)}}{h^2}+\frac{{c^{(0)}}'}{2}\right)(\Phi_I^{(1)})^\prime
+\frac{J^{2/3}}{r^2h}\left(\frac{ 1}{r^2}(2\mu \alpha^{(0)} -\mu^2c^{(0)})\right)\Phi_I^{(1)}  \nonumber \\ &
-\frac{J^{2/3} b^{(0)}}{r^2h^2}\left(\frac{\mu^2}{r^2} -m^2\right)\Phi_I^{(1)} = -  \frac{2}{3r^4h} \mu J^{-2/3} \dot J  \chi_{R}^{(0)}.
\end{align}
As in the previous section, the source is proportional to the time derivative of $J(t)$ which arises from the time derivative of $\Phi_R^{(0)}$. Since the operator $D$ has a zero mode we can estimate $\Phi_I^{(1)}$ by perturbation theory, leading to
\ben
\Phi_I^{(1)} \sim J^{-4/3}{\dot J},
\een
as in (\ref{2-3-6}). The condition for adiabaticity breakdown is therefore the same as in section 2, equation (\ref{2-3-8}).

The other equation which contains time derivatives is the $(rt)$ component of the Einstein equations,  which becomes
\begin{align}
-\frac{3 \dot B}{2rB} - \frac{1}{4} \dot A A' - & \frac{\dot B A'}{2B}- \frac{\dot A B'}{4B}+ \frac{\dot A C'}{4}
+ \frac{\dot B C'}{4B}- \frac{1}{2} \dot A' - \frac{\dot B'}{2B} \nonumber \\
&- \dot \Phi_R \Phi'_R - \dot \Phi_I \Phi'_I-\phi \Phi_I \Phi'_R  + \phi \Phi_R \Phi_I' =0.
 \label{constraint_grt}
\end{align}
This is a constraint equation and
once this equation is satisfied at some constant $r$ slice, the $(tt)$ component of Einstein equations guarantee that this is satisfied everywhere. For large values of $r$ the equation (\ref{constraint_grt}) becomes
\begin{align}
\dot {\cal T}_{tt} = 2 \mu  B_{I}^{(1)} A_{R}^{(0)}+ O(J^{1/3} \dot J)
\label{energy-conservation},
\end{align}
where we have written $\Phi_{I,R}^{(a)} \sim A_{I,R}^{(a)} r^{-\Delta_-} +
B_{I,R}^{(a)} r^{-\Delta_+}$ and ${\cal T}_{tt}$ denotes the boundary energy momentum tensor
\ben
{\cal T}_{tt} = \frac{1}{8\pi G}\left( -\frac{1}{2} +\frac{1}{2}J^{2/3}h_{tt} -2 J^{2/3} h_{rr} \right),
\een
and the asymptotic form of the metric components can be shown to be
\begin{align}
g_{tt} & \rightarrow r^2 (-1-\frac{J^{2/3}h_{tt}}{r^4}), \nonumber \\
g_{rr} & \rightarrow \frac{1}{r^2 h(r)} \left(1- \frac{J^{2/3}h_{rr}}{r^4} \right)  .
\end{align}
Clearly both sides of (\ref{energy-conservation}) are proportional to $J^{-1/3}{\dot{J}}$. This shows the consistency of our lowest order adiabatic solution.

\subsection{Scaling Solution}

In the critical region we may set $J(t) \sim J_0 vt$. Following the analysis in section \ref{sec-scaling}, we now show that the complete set of equations of motion have scaling solutions. As in section \ref{sec-scaling}, the first step is to rescale time
\begin{align}
t \to \eta = v^{2/5} t,
\end{align}
and separate out the source part and fields as
\begin{align}
\Phi_R = J(v^{-2/5}\eta ) r^{-\Delta_-} + v^{1/5} \chi_R, \quad 
\Phi_I = v^{1/5} \chi_I, \quad \phi = \mu + v^{2/5} \alpha, \nonumber \\
 \quad A= v^{2/5} a, \quad B= h + v^{2/5} b, \quad C= v^{2/5} c.
\end{align}
The equations of motion can be now expanded for small $v$. The scalar field equations (\ref{eom_phi_R_gr_full}) and (\ref{eom_phi_I_gr_full}) lead to 
\begin{align}
&
v^{1/5}D(\chi_R)+
v^{3/5 }\Biggl[ J_0 \eta D(r^{-\Delta_-})+
\left(\frac{a'}{2}+\frac{b'}{h}-\frac{h'b'}{h^2}+\frac{c'}{2}\right)\chi'_R
 \nonumber \\
&
+
\frac{1}{r^2h}
\left(
-\frac{b}{h}\left(\frac{\mu^2}{r^2}-m^2\right)
-\frac{\mu^2c}{r^2}
+\frac{2\mu \alpha}{r^2}
\right)
\chi_R
-\frac{2\mu}{r^4h}  \dot \chi_I \Biggr]+ O(v)=0 ,
\label{scaling-gr-phi_R}
\\
&
v^{1/5}D(\chi_I)+
v^{3/5 }\Biggl[
\left(\frac{a'}{2}+\frac{b'}{h}-\frac{h'b'}{h^2}+\frac{c'}{2}\right)\chi'_I
 \nonumber \\
&
+
\frac{1}{r^2h}
\left(
-\frac{b}{h}\left(\frac{\mu^2}{r^2}-m^2\right)
-\frac{\mu^2c}{r^2}
+\frac{2\mu \alpha}{r^2}
\right)
\chi_I
+\frac{2\mu}{r^4 h}  \dot \chi_R \Biggr]+ O(v)=0 .
\label{scaling-gr-phi_I}
\end{align}
The Maxwell equation (\ref{eom_gauge_gr_full}) yield
\ben
v^{2/5} \left[ \frac{1}{r^3} \partial_r \left(r^3 h \partial_r \alpha \right) - \frac{2\mu}{r^2} \left(\chi_R^2+\chi_I^2 \right) \right]+O(v^{4/5}) =0 ,
\label{scaling-gr-gauge}
\een
while the Einstein equations (\ref{eom_A_gr_full}) - (\ref{eom_B_gr_full}) become
\begin{align}
&v^{2/5}\left[a'-\frac{r^2c''+2rc'+2r^2\left( \chi'^2_R+\chi'^2_I \right)}{3r}\right]+ O(v^{4/5}) =0\;, \\
&v^{2/5}\left[c''+\left(\frac{5}{r}+\frac{h'}{h}\right)c'-\frac{2\mu^2\left(\chi_R^2+\chi_I^2\right)}{r^4h}\right]+O(v^{4/5})=0,\\
&v^{2/5}\left[\frac{3}{r}b'-\frac{h'}{2}c'+\frac{12}{r^2}b+h\left(\chi'^2_R+\chi'^2_I\right) 
+\frac{1}{r^2}\left(\frac{\mu^2}{r^2}+m^2\right)\left(\chi_R^2+\chi_I^2\right)\right]+O(v^{4/5})=0\;.
\label{scaling-gr-gravity}
\end{align}
Here we have omitted the constraint equations.
We see that the gauge and gravity fields $\alpha$, $a$, $b$ and $c$ are solved by using the Green's function for the operators which appear in (\ref{scaling-gr-gauge}) - (\ref{scaling-gr-gravity}) in a way similar to equation (\ref{2-4-8}).
It is clear from the scalar field equations that the zero mode of $D$ dominates in the equation (\ref{scaling-gr-phi_R}) and (\ref{scaling-gr-phi_I}).
Therefore, in a way entirely analogous to section \ref{sec-scaling}, we obtain the scaling relations
\begin{align}
\langle \cO (t,v) \rangle & = v^{1/5} F(tv^{2/5}), \\
\langle \rho (t,v) \rangle& = v^{2/5} G(tv^{2/5}), \\
\langle T_{\mu\nu} (t,v) \rangle& = v^{2/5} H_{\mu\nu}(tv^{2/5}), 
\end{align}
where we have used the usual identification of the subleading pieces of the bulk fields for $r \rightarrow \infty$ with the expectation values of the dual operators.

\section{Conclusions and Discussions} \label{sec5}

This work demonstrates that the mechanism for emergence of Kibble-Zurek scaling in holographic models found in \cite{basu-das} - \cite{bdds} is robust in the sense that it continues to hold when one includes the backreaction of the gauge field and the metric. While we have shown this in a model of a holographic superfluid, we expect that this will hold for other models of critical points, e.g. the model with double trace deformations \cite{doubletrace} studied in \cite{bdds}. We have restricted our attention to the nature of the solution in the critical region $ t \sim 0$ and concentrated on the emergence of scaling behavior.

As mentioned above, the late time behavior of the dynamics could be interesting.
If we performed the slow quench far away from a critical point, one expects the response is adiabatic and gravitational collapse does not occur. The breakdown of adiabaticity in the gauge and gravity sector implies that at late times the  background will change substantially. It is important to determine if the late time state is a steady state and if there is thermalization. We leave this problem, which requires serious numerical work, for the future.

In this paper we have considered global quantum quench in holographic models in the limit where the bulk description is purely classical. In the field theory this means we are considering the leading term in the $N=\infty$ limit, where fluctuations of gauge invariant observables are suppressed. Consequently the solutions and the resulting order parameter are spatially homogeneous (in the field theory space directions). This limit is adequate to uncover the {\em scaling properties} of local observables, but not adequate to discuss the other important aspect of Kibble-Zurek physics, viz. defect formation, which requires inhomogeneous solutions. The latter requires a treatment of fluctuations, which have been modelled by adding random noise to the bulk equations in \cite{Sonner:2014tca}. It would be interesting to see what happens to the route to scaling in the presence of such noise.

\section{Acknowledgements}

We would like to thank Pallab Basu, Diptarka Das, Juan Maldacena,
 Tadashi Takayanagi and Toby Wiseman for discussions.
 We would also like to thank M. Headrick for his Mathematica package for performing the calculations in section \ref{sec4}.
T. M. would like to thank Yukawa Institute for hospitality where part of the work was done. This work is partially supported by National Science Foundation grant PHY-0970069 and a Visiting Professorship at Yukawa Institute for Theoretical Physics.

\appendix

\section{Complete equations of motion in section \ref{sec4}}
\setcounter{equation}{0}

In this appendix, we show the equations of motion (\ref{eom-scalar}) - (\ref{eom-gr}) explicitly.
The equations for the scalars become
\begin{align}
&\Phi_R''+\left(\frac{5}{r}+\frac{A'}{2}+\frac{B'}{B}+\frac{C'}{2}\right)\Phi_R'+\frac{1}{r^2B}\left(\frac{e^{-C}\phi^2}{r^2}-m^2\right)\Phi_R \nonumber \\
&-\frac{e^{-C}}{r^4 B}\left[
\ddot \Phi_R + \frac{\dot A- \dot C}{2} \dot \Phi_R
+2\phi \dot \Phi_I + \dot \phi \Phi_I+ \frac{\phi}{2}(\dot A- \dot C)\Phi_I
\right]
=0\;,
\label{eom_phi_R_gr_full}  \\
&\Phi_I''+\left(\frac{5}{r}+\frac{A'}{2}+\frac{B'}{B}+\frac{C'}{2}\right)\Phi_I'+\frac{1}{r^2B}\left(\frac{e^{-C}\phi^2}{r^2}-m^2\right)\Phi_I \nonumber \\
&-\frac{e^{-C}}{r^4 B}\left[
\ddot \Phi_I + \frac{\dot A- \dot C}{2} \dot \Phi_I
-2\phi \dot \Phi_R - \dot \phi \Phi_R- \frac{\dot A- \dot C}{2}\phi\Phi_R
\right]
=0\;.
\label{eom_phi_I_gr_full}
\end{align}
The Maxwell equation becomes 
\begin{equation}
\phi''+\left(\frac{3}{r}+\frac{A'}{2}+\frac{B'}{B}-\frac{C'}{2}\right)\phi'-\frac{2}{r^2B}\left( \dot \Phi_R \Phi_I- \Phi_R \dot \Phi_I +   |\phi\Phi^2| \right)=0\;.
\label{eom_gauge_gr_full}
\end{equation}
We also have the constraint equation corresponding to (\ref{eom_ar_ads}) but we omit it here.

Following \cite{horo}, we take linear combinations of the Einstein equations.
From $g^{\theta\theta}(G_{\theta\theta}-T_{\theta\theta})-g^{rr}(G_{rr}-T_{rr})=0$ (where $G_{AB}$ is the Einstein tensor and $T_{AB}$ is the bulk energy momentum tensor),  we obtain 
\begin{align}
A'=&\frac{2r^2C''+r^2C'^2+4rC'+4r^2\left(\Phi_R'^2+\Phi_I'^2\right)-2e^{-C}\phi'^2}{r(6+rC')} \nonumber \\
&+\frac{e^{-C}}{r^3B(6+rC')}\left[
2 \ddot A+ \dot A^2 + \frac{2 \dot A \dot B}{B} + \frac{4 \ddot B}{B} - \frac{4 \dot B^2}{B^2}
 - \dot A \dot C - \frac{2 \dot B \dot C}{B}
\right],
\label{eom_A_gr_full}
\end{align}
and, from $g^{xx}(G_{xx}-T_{xx})-g^{tt}(G_{tt}-T_{tt})=0$, we obtain

\begin{align}
&C''+\frac12 C'^2+\left(\frac{5}{r}+\frac{A'}{2}+\frac{B'}{B}\right)C'-\left(\phi'^2+\frac{2\phi^2(\Phi^2_R+\Phi^2_I)}{r^2B}\right)\frac{e^{-C}}{r^2} \nonumber \\
& +\frac{e^{-C}}{r^4 B}\left[- \ddot A
-\frac{\dot A^2}{2}- \frac{ \dot A \dot B }{B}-  \frac{\dot B^2}{B^2}
+ \frac{\dot A \dot C}{2} + 4 \phi \left( \Phi_R \dot \Phi_I - \dot \Phi_R  \Phi_I \right)
-2 \left(\dot \Phi_R^2+ \dot \Phi_I^2 \right)
 \right]
=0.
\label{eom_C_gr_full}
\end{align}
From $g^{xx}(G_{xx}-T_{xx})-g^{tt}(G_{tt}-T_{tt})-g^{\theta\theta}(G_{\theta\theta}-T_{\theta\theta})=0$, we obtain
\begin{align}
&B'\left(\frac{3}{r}-\frac{C'}{2}\right)+B\left(\Phi_R'^2+\Phi_I'^2-\frac12 A'C'+\frac{e^{-C}\phi'^2}{2r^2}+\frac{12}{r^2}\right) \nonumber \\
&+\frac{1}{r^2}\left(\frac{e^{-C}\phi^2(\Phi^2_R+\Phi^2_I)}{r^2}+m^2(\Phi^2_R+\Phi^2_I)-12\right) \nonumber \\
&+\frac{e^{-C}}{r^4 }\left[ \ddot A
+\frac{\dot A^2}{2}+ \frac{ \dot A \dot B }{B} +\frac{\ddot B}{B}-  \frac{\dot B^2}{2B}
- \frac{\dot A \dot C}{2}- \frac{\dot B \dot C}{2}
 -2 \phi \left( \Phi_R \dot \Phi_I - \dot \Phi_R  \Phi_I \right)
+ \left(\dot \Phi_R^2+ \dot \Phi_I^2 \right)
 \right]
=0\;.
\label{eom_B_gr_full}
\end{align}
The (tr) component of the equation of motion, which is a constraint equation, becomes
\begin{align}
&-\frac{3 \dot B}{2rB} - \frac{1}{4} \dot A A' - \frac{\dot B A'}{2B}- \frac{\dot A B'}{4B}+ \frac{\dot A C'}{4}
+ \frac{\dot B C'}{4B}- \frac{1}{2} \dot A' - \frac{\dot B'}{2B} \nonumber \\
&- \dot \Phi_R \Phi'_R - \dot \Phi_I \Phi'_I+\phi \left(  \Phi_R \Phi_I' - \Phi_I \Phi'_R  \right) =0.
 \label{constraint_grt_full}
\end{align}

\end{document}